# Entanglement between two atoms in a damping Jaynes-Cummings model

Guo-Feng Zhang(张国锋)[1,2*], Xin-Chen Xie(谢心澄)[2]

1 Department of Physics, School of Physics and Nuclear Energy Engineering,

Beijing University of Aeronautics and Astronautics,

Xueyuan Road No. 37, Beijing 100191, People's Republic of China

2 Department of Physics, Oklahoma State University, Stillwater, OK 74078

## Abstract

The entanglement between two atoms in a damping Jaynes-Cummings model is investigated with different decay coefficients of the atoms from the upper level to other levels under detuning between the atomic frequency and the quantized light field frequency. The results indicate that the larger the decay coefficient is, the more quickly the entanglement decays. The detuning enhances the entanglement's average value at long times. More importantly, the results show that the so-called sudden death effect can be avoided by enhancing the detuning or the decay coefficient.

* Corresponding author. E-mail: gf1978zhang@buaa.edu.cn

Entanglement is one of the most bizarre features that cannot be found in the classical world. It plays a fundamental role in almost all efficient protocols for quantum computation (QC) and quantum information processing (QIP). In recent years, there has been an ongoing effort to characterize qualitatively and quantitatively the entanglement's properties and apply them to quantum communication and information. Among the studies of entanglement, the dynamic evolution and control appear very important. Recently, the dynamics of entanglement in bipartite systems has received renewed attention since the work of Yu and Eberly [1], in which the entanglement between two particles coupled to two independent environments completely vanished in a finite time. This surprising phenomenon, termed the sudden death effect (SDE), contrary to intuition based on experience of qubit decoherence, has stimulated great interest [2–7]. In Ref. [3, 4], the authors show that for a special initial state the entanglement disappears in a finite time and is then revived after a dark period because of the interaction between the particles, which is different from the works by Yu and Eberly [1] since in Ref. [3, 4] the effects of interaction between the particles and the couplings to the same environment have been discussed extensively. In Ref.[5], the authors investigate the entanglement dynamics of two isolated atoms, each in its own lossless Jaynes-Cummings (JC) cavity. They show the existence of the so-called sudden death effect (SDE) for the resonant case. In Ref. [8] and [9], the entanglement dynamics in a double JC model and a four-qubit model for the non-resonant case are investigated, respectively, when the cavities are lossless. Moreover, two-qubit states with a desired degree of entanglement can be generated by varying the atom-cavity coupling [10]. The varying atom-field coupling can change the dynamic properties of the atom and the field with one photon transiations greatly [11]. However, these results are obtained for the case in which the decay of the upper level is ignored. Considering the experimental realization, we study the JC model including dissipation of the upper level in the resonant and non-resonant cases. We will be the first to consider the effect of entanglement decay of the one-photon state to other photon states in the double JC model. The results indicate that the dissipation of the upper level leads to the decay and the reduction of the SDE of the entanglement. Detuning modulation can control the entanglement properties, moreover, SDE can be avoided by applying high detuning and decay coefficient.

The two atoms are in a perfect one-mode cavity with losses, but there is no interaction between the two atoms. We can get a mixed state of atom $A$ and atom $B$ by tracing over the cavity modes at time $t$. For a pair of qubits, we will adopt concurrence as the entanglement measure ($1 \geq C \geq 0$).

The Hamiltonian of the system within the rotating wave approximation (RWA) can be written as [12–14]

$$H = \omega\sigma_z^A + \omega\sigma_z^B + \nu a^\dagger a + \nu b^\dagger b + g(a^\dagger \sigma_-^A + a\sigma_+^A) + g(b^\dagger \sigma_-^B + b\sigma_+^B) - i\frac{\gamma}{2}(|e\rangle_B\langle e| + |e\rangle_A\langle e|)), \quad (1)$$

where $a(b)$ and $a^\dagger(b^\dagger)$ denote the annihilation and creation operators for the two cavity fields respectively, and $\sigma_z^l = |e\rangle_l\langle e| - |g\rangle_l\langle g|$, $\sigma_+^l = |e\rangle_l\langle g|$, $\sigma_-^l = |g\rangle_l\langle e| (l = A, B)$ are the atomic operators. We assume that the transition frequencies of atom $A$ and atom $B$ are $\omega$, both coupling coefficients between the atoms and the fields are $g$. $\gamma$ is the decay coefficient from the upper level $|1\rangle$ of the atoms to other levels, i.e. the dissipation coefficient. Clearly, there will be no interaction between atom $A$ and atom $B$ or between cavity $a$ and cavity $b$.

For greatest simplicity, we assume that the cavities are prepared initially in the vacuum state $|0_a\rangle \otimes |0_b\rangle = |00\rangle$ and that the two atoms are in a pure entangled state specified below. Under these assumptions, there is never more than one photon in each cavity. This allows a uniform measure of quantum entanglement-concurrence. In principle, there are six different concurrences that provide information about the overall entanglements that may arise. Here we confine our attention to the concurrence between the two atoms $C_{AB}$.

We assume that the two atoms are initially in a partially entangled atomic pure state that is a combination of Bell states usually denoted $|\Psi^\pm\rangle$, we have

$$|\Psi_{atom}\rangle = \cos[\alpha]|eg\rangle + \sin[\alpha]|ge\rangle, \quad (2)$$

with the first index denoting the state of atom $A$ and the second denoting the state of atom $B$ ($|g\rangle$ is the ground state and $|e\rangle$ is the excited state). Thus the initial state for the system (1) can be given by

$$|\Psi(0)\rangle = |\Psi_{atom}(0)\rangle \otimes |00\rangle = \cos[\alpha]|eg00\rangle + \sin[\alpha]|ge00\rangle. \quad (3)$$

The solution of the system in terms of the standard basis can be written as:

$$|\Psi(T)\rangle = x_1[T]|eg00\rangle + x_2[T]|ge00\rangle + x_3[T]|gg10\rangle + x_4[T]|gg01\rangle, \quad (4)$$

with the corresponding coefficients given by the following:

$$\begin{aligned}
x_1[T] &= \frac{1}{2} M_- \cos[\alpha] e^{-\frac{T(\xi_+ + \eta_-)}{4}}, \\
x_2[T] &= \frac{1}{2} M_- \sin[\alpha] e^{-\frac{T(\xi_+ + \eta_-)}{4}}, \\
x_3[T] &= -2i \cos[\alpha] \frac{\zeta_-}{\eta_-} e^{-\frac{T(\xi_+ + \eta_-)}{4}}, \\
x_4[T] &= -2i \sin[\alpha] \frac{\zeta_-}{\eta_-} e^{-\frac{T(\xi_+ + \eta_-)}{4}},
\end{aligned} \qquad (5)$$

with $T = gt$, $\delta = \nu - \omega$ is the detuning, $\kappa = \gamma/g$ and $\xi_\pm = \kappa \pm (2i\delta/g)$, $\eta_\pm = \sqrt{-16 + \xi_\pm}$. $\zeta_\pm = -1 + e^{T\eta_\pm/2}$, $M_\pm = 1 + e^{T\eta_\pm/2} - \zeta_\pm \xi_\pm/\eta_\pm$. The information about the entanglement of the two atoms is contained in the reduced density matrix $\rho_{AB}$ which can be obtained from Eq.(4) by tracing out the photonic part of the total pure state. In the standard basis $\{|e,e\rangle, |e,g\rangle, |g,e\rangle, |g,g\rangle\}$, the density matrix $\rho_{AB}$ can be expressed as

$$\rho_{AB} = \begin{pmatrix} 0 & 0 & 0 & 0 \\ 0 & |x_1|^2 & x_1 x_2^* & 0 \\ 0 & x_2 x_1^* & |x_2|^2 & 0 \\ 0 & 0 & 0 & |x_3|^2 + |x_4|^2 \end{pmatrix}, \qquad (6)$$

where $*$ stands for the complex conjugate. It can be shown that the concurrence of the density matrix (6) is given by

$$C_{AB}(T) = 2 \max\{|x_1 x_2^*|, 0\} = \max\{\frac{1}{4} e^{-T(\xi_+ + \xi_- + \eta_+ + \eta_-)/4} M_+ M_- \sin[2\alpha], 0\}. \qquad (7)$$

Now we assume that the initial state for the total system is in a combination of the other two Bell states $|\Phi^\pm\rangle$

$$|\Phi(0)\rangle = \cos[\alpha]|ee00\rangle + \sin[\alpha]|gg00\rangle, \qquad (8)$$

in which case the state of the total system at time $t$ can be expressed in the standard basis as

$$|\Phi(T)\rangle = x_1[T]|ee00\rangle + x_2[T]|gg11\rangle + x_3[T]|eg01\rangle + x_4[T]|ge10\rangle + x_5[T]|gg00\rangle, \qquad (9)$$

where the coefficients are now given by

$$x_1[T] = \frac{1}{2\eta_-^2}\{\cos[\alpha]e^{-T(\kappa+2i\nu/g+\eta_-)/2}[-8(2+\zeta_-)^2 - \zeta_-(2+\zeta_-)\eta_-\xi_- + (1+(1+\zeta_-)^2)\zeta_-^2]\},$$

$$x_2[T] = -4e^{-T(\kappa+2i\nu/g+\eta_-)/2}(\frac{\zeta_-}{\eta_-})^2\cos[\alpha],$$

$$x_3[T] = x_4[T] = -\frac{i}{\eta_-^2}\{\cos[\alpha]e^{-T(\kappa+2i\nu/g+\eta_-)/2}[-\zeta_-^2\xi_- + (-1+e^{T\eta_-})\eta_-]\},$$

$$x_5[T] = e^{2iT\omega/g}\sin[\alpha]. \qquad (10)$$

In the basis of $|e,e\rangle$, $|e,g\rangle$, $|g,e\rangle$ and $|g,g\rangle$ the reduced density matrix $\rho_{AB}$ is now found to be

$$\rho_{AB} = \begin{pmatrix} |x_3|^2 & 0 & 0 & 0 \\ 0 & |x_1|^2 & x_1 x_5^* & 0 \\ 0 & x_5 x_1^* & |x_2|^2 + |x_5|^2 & 0 \\ 0 & 0 & 0 & |x_4|^2 \end{pmatrix}, \qquad (11)$$

and the concurrence for this matrix is given by $C_{AB}(T) = \max\{F(T) + G(T), 0\}$, with $F(T) = 32e^{-\kappa T}\cos^2[\alpha]\sinh[\frac{T\eta_+}{4}]\sinh[\frac{T\eta_-}{4}]\Delta_+\Delta_-/(\eta_+\eta_-)^2$ and $G(T) = 2\sqrt{\Lambda_+\Lambda_-}\sin^2[\alpha]\cos^2[\alpha]/(\eta_+\eta_-)^2 e^{-T\kappa/2}$, where $\Delta_\pm = \mp\cosh[T\eta_\pm/4]\eta_\pm \pm \sinh[T\eta_\pm/4]\xi_\pm$ and $\Lambda_\pm = 8 + \sinh[T\eta_\pm/2]\eta_\pm\xi_\pm - \cosh[T\eta_\pm/2](-8+\xi_\pm^2)$.

In the following, we will study entanglement properties based on two different kinds of initial states for both resonant and non-resonant cases without and with a decay coefficient.

**RESONANT CASE**

We first consider the time evolution of the entanglement in the resonant case, i.e. $\delta = \nu - \omega = 0$. For the first initial state, the entanglement of atom $A$ and $B$ becomes $C_{AB} = \max\{f(T)/4, 0\}$, with

$$f(T) = \exp[-\frac{T(\kappa+\eta)}{2}]\{1 + \frac{\kappa}{\eta} + \exp[\frac{T\eta}{2}](1-\frac{\kappa}{\eta})\}^2\sin[2\alpha], \qquad (12)$$

and $\eta = \sqrt{-16+\kappa^2}$. From Eq.(12), we can find that $C_{AB}$ cannot reach zero so long as $T$ is finite and remain zero for a non-zero time segment, i.e., there is no SDE, as shown in Fig. 1. Here we term cases $\gamma = 0$, $\gamma = 0.5g$ and $\gamma = g$ as zero dissipation, weak dissipation and strong dissipation respectively. After the beginning of interaction ($T > 0$), the concurrences return to their minimum values gradually, after that, they turn back and oscillate regularly with the same period. In the actual system, the atom is steady when it is in the ground state $|g\rangle$. However, when the atom is in the excited state $|e\rangle$ the factors such as the spontaneous emission will lead to decay of the upper level atom. With the increase of $\gamma$, the value of concurrence that can arise becomes smaller and the concurrence tends to zero rapidly which can be seen from Fig. 1. Due to the background dissipation, the atom in the upper level will partly decay to other levels apart from the ground state. Hence the decay coefficient will influence the decay rate of the concurrence. However, the SDE still will not appear. For the second initial state, unlike the pervious case, Fig. 2 shows that the concurrence can fall abruptly to zero for some $\alpha$ values and will remain zero for a period of time before recovering. Obviously, the length of the time interval for the zero concurrence is dependent on the degree of entanglement of the initial state. The smaller the initial degree of concurrence, the longer the state will stay in the disentangled separable state. These conclusions are the same as those in Ref.[5]. Comparing Fig. 2 (a), (b) and (c), one can find that the decay coefficient $\gamma$ can weaken the SDE. The length of the time interval for the zero concurrence can be shortened by introducing larger $\gamma$, however, the value of concurrence which can be reached will become smaller and the concurrence will tend to zero more rapidly. In order to see how the decay coefficient influences SDE, Fig. 3 give an illustration for $\alpha = \pi/6$. As can be seen, for the $\gamma = g$ case, there is no SDE; but for the $\gamma = 0$ case, SDE will be prominent.

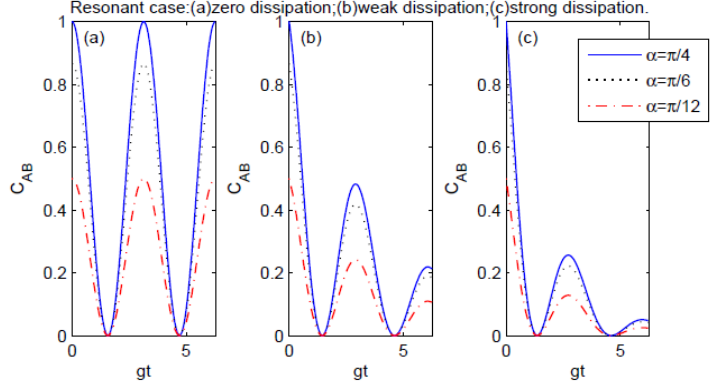

FIG. 1: (Colour online) The concurrence for atom-atom entanglement with the initial state $|\Psi_{atom}\rangle = \cos[\alpha]|eg\rangle + \sin[\alpha]|ge\rangle$ for zero detuning $\delta = \nu - \omega = 0$. (a) corresponds to the case in which the decay coefficient $\gamma$ from the upper level of the atom to the other levels is 0, (b) corresponds to $\gamma = 0.5g$, (c) corresponds to $\gamma = g$.

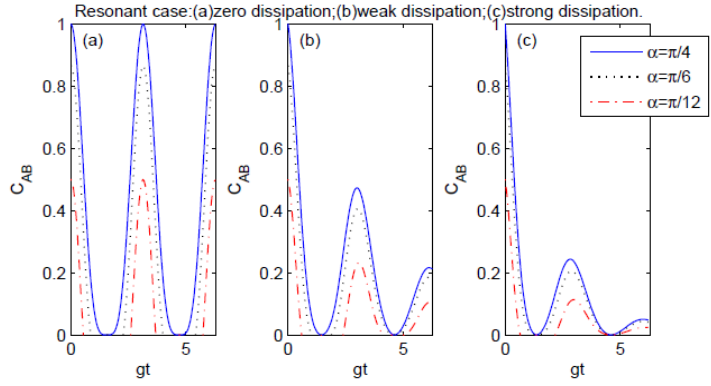

FIG. 2: (Colour online) The concurrence for atom-atom entanglement with the initial state $|\Psi_{atom}\rangle = \cos[\alpha]|ee\rangle + \sin[\alpha]|gg\rangle$ for zero detuning $\delta = \nu - \omega = 0$. (a) corresponds to the case in which the decay coefficient $\gamma$ from the upper level of the atom to the other levels is 0, (b) corresponds to $\gamma = 0.5g$, (c)corresponds to $\gamma = g$.

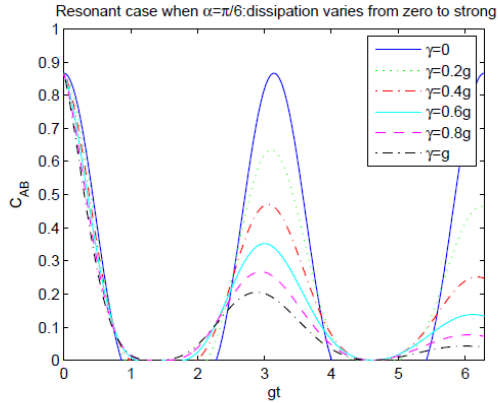

FIG. 3: (Colour online) The concurrence for atom-atom entanglement with the initial state $|\Psi_{atom}\rangle = \cos[\alpha]|ee\rangle + \sin[\alpha]|gg\rangle$ for $\alpha = \pi/6$.

**NON-RESONANT CASE**

In the following, we extend our work to the non-resonant case. We will show that the detuning modulation can control the entanglement properties, even change the entanglement evolution behaviour, as shown in Fig. 4 and Fig. 5. The detuning of the atom and the field frequencies enhances the entanglement in some short time regions. However, it can not prolong the entanglement existence time. The concurrence can be larger when a non-zero detuning is introduced when $\gamma = 0$. The evolution period becomes shorter with increasing of detuning. For the second initial state, increasing detuning can weaken SDE. When $\gamma = g$, the concurrence evolves in a very similar way for the two different initial states. The above conclusions can also be seen from Fig.6 and Fig.7 where one can find that the concurrence is symmetric with respect to the detuning.

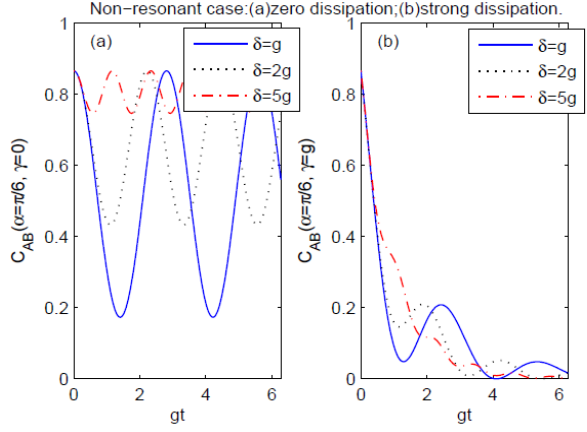

FIG. 4: (Colour online) The concurrence for atom-atom entanglement with the initial state $|\Psi_{atom}\rangle = \cos[\alpha]|eg\rangle + \sin[\alpha]|ge\rangle$ and $\delta = \nu - \omega$.

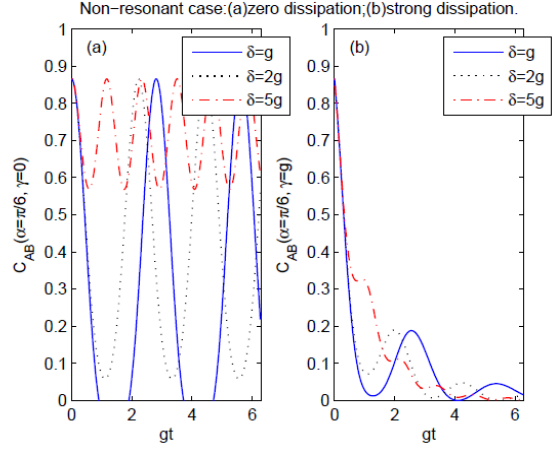

FIG. 5: (Colour online) The concurrence for atom-atom entanglement with the initial state $|\Psi_{atom}\rangle = \cos[\alpha]|ee\rangle + \sin[\alpha]|gg\rangle$ and $\delta = \nu - \omega$.

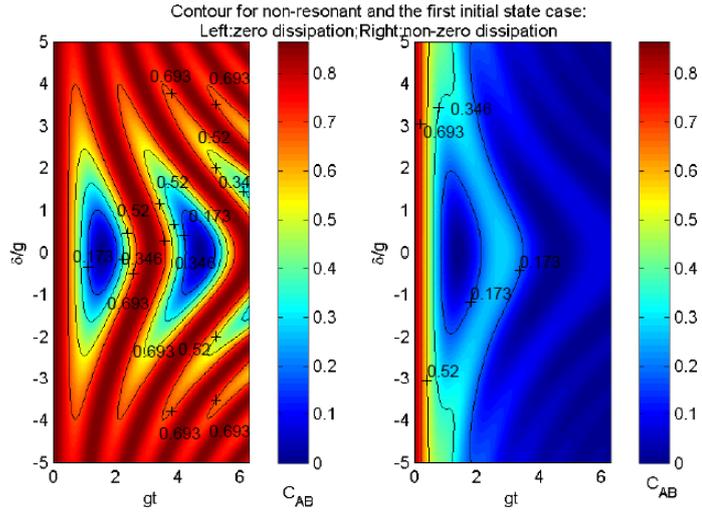

FIG. 6: (Colour online) The contour of concurrence for atom-atom entanglement with the initial state $|\Psi_{atom}\rangle = \cos[\alpha]|eg\rangle + \sin[\alpha]|ge\rangle$ for $\alpha = \pi/6$ and $\delta = \nu - \omega$. Left:$\gamma = 0$, right:$\gamma = 0.8g$.

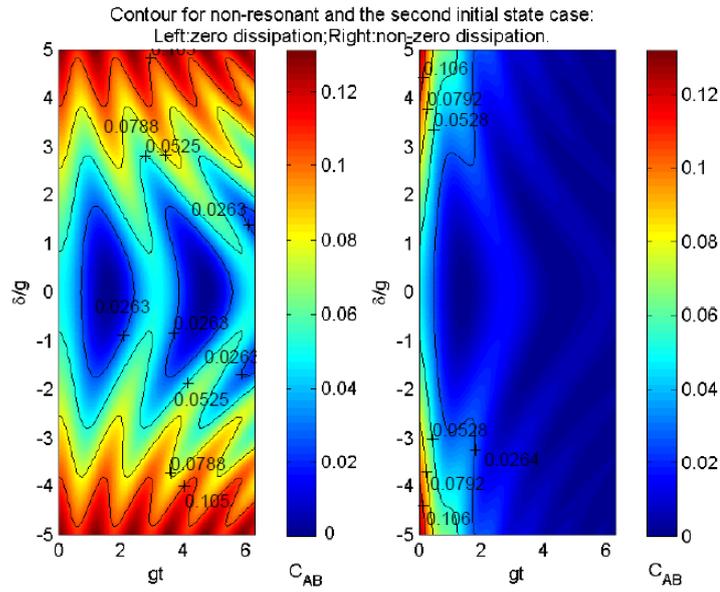

FIG. 7: (Colour online) The contour of concurrence for atom-atom entanglement with the initial state $|\Psi_{atom}\rangle = \cos[\alpha]|ee\rangle + \sin[\alpha]|gg\rangle$ for $\alpha = \pi/6$ and $\delta = \nu - \omega$. Left:$\gamma = 0$, right:$\gamma = 0.8g$.

In summary, we have investigated the entanglement between two atoms in two independent damping Jaynes-Cummings models for both resonant and non-resonant cases. We found that the detuning between cavity frequency and atomic frequency can increase the maximum value of the entanglement. Moreover, it can weaken SDE when the two atoms' initial state is $|\Psi_{atom}\rangle = \cos[\alpha]|ee\rangle + \sin[\alpha]|gg\rangle$. The decay coefficient plays a very important role in reviving the entanglement. The length of the time interval for the zero concurrence can be shortened by introducing a larger decay coefficient. However, the value that entanglement can reach will become smaller and it will tend to zero more rapidly. For the initial atomic state $|\Psi_{atom}\rangle = \cos[\alpha]|eg\rangle + \sin[\alpha]|ge\rangle$, there is no SDE regardless of the value of $\gamma$.

**ACKNOWLEDGEMENTS**

This work was supported by the National Science Foundation of China under Grants No. 10874013 and 10604053.